\documentclass[aps,amsmath,twocolumn,preprintnumbers,floatfix,nofootinbib,raggedfooter,twoside,showpacs,showkeys]{revtex4}
\usepackage{psfrag,alltt,graphicx}
\usepackage{subfigure}
\usepackage[T1]{fontenc} 
\usepackage{textcomp}             % render \textregistered nicely
\usepackage[breaklinks]{hyperref} % ensure hyperref is loaded before breakurl
\usepackage{breakurl}             % redefine \url to allow breaks in postscript
\newcommand{\graphicsdir}{pics/}
\renewcommand{\emph}[1]{\textsl{#1}}
\newcommand{\pack}[1]{\NoCaseChange{\textsf{#1}}}    % package name
\newcommand{\mc}[1]{{\texttt{#1}}}                   % Mathematica command
\newcommand{\acro}[1]{{\small #1}}                   % acronym
\providecommand{\cs}[1]{%
   \texttt{\expandafter\string\csname #1\endcsname}} % TeX macro
\providecommand{\latex}[1]{\texttt{#1}}              % dito
\providecommand{\env}[1]{\cs{begin{#1}}}             % LaTeX environment
\providecommand{\marg}[1]{$\langle$\texttt{\textsl{#1}}$\rangle$} % mandatory argument
\providecommand{\oarg}[1]{$[$\texttt{\textsl{#1}}$]$}             % optional -"-
\newcommand{\Rule}{\(\to\)\allowbreak}                          % -> for textmode
\newcommand{\amsmath}{\texttt{amsmath}}              
\newcommand{\Mathematica}{\NoCaseChange{\textsc{Mathematica}}}      %${}^\text{\textregistered}$}
\newcommand{\MathematicaR}{\Mathematica\textsuperscript{\tiny\textregistered}}
\newcommand{\PSfrag}{\pack{PSfrag}}
\newcommand{\MathPSfrag}{\pack{MathPSfrag}}
\begin{document}
\title{\MathPSfrag: Creating Publication-Quality Labels in \Mathematica\ Plots}
\author{Johannes  Gro{\ss}e$^*$}
%\thanks{jgrosse@mppmu.mpg.de}
\affiliation{Max-Planck-Institut f{\"u}r Physik (Werner-Heisenberg-Insitut),\\ 
   F{\"o}hringer Ring 6, 80805 M{\"u}nchen, Germany}

\affiliation{\vspace{0.5ex}
   Arnold-Sommerfeld-Center for Theoretical Physics, Department f{\"u}r Physik,
   Ludwig-Maximilians-Universit{\"a}t M{\"u}nchen, Theresienstra{\ss}e 73, 80333 M{\"u}nchen, Germany
}
%\email{jgrosse@mppmu.mpg.de}
%\homepage{http://wwwth.mppmu.mpg.de/members/jgrosse}
\date{October 30, 2005}

\begin{abstract}
This article introduces a \MathematicaR{} package providing
a graphics export function that \emph{automatically} replaces
\MathematicaR{} expressions in a graphic by the corresponding 
\LaTeX\ constructs and positions them correctly. 
It thus facilitates the creation of publication-quality Enscapulated
PostScript (\acro{EPS}) graphics.
\end{abstract}

 % Surveys and tutorial papers; resource letters
%\textsl{1991 MSC:} 68U10 % Image processing
\preprint{LMU-ASC 70/05}
\preprint{MPP-2005-126}
\preprint{cs.GR/0510087}
\maketitle

\section*{Introduction}

Many programs producing \acro{EPS} graphics do not allow the inclusion
of \LaTeX\ commands. While there exist several solutions (see for example
\cite{McKay:1999}) to work
around these difficulties, they all have various drawbacks.  In this
article, we will focus on a particular existing solution, the
\PSfrag{} package \cite{Grant:1998}, which provides \LaTeX\ macros
allowing to replace pieces of text (``tags'') in an \acro{EPS} file by
an arbitrary \LaTeX\ construct.

However, for \PSfrag{} to work, the application must write tags unaltered
into the \acro{EPS} file. For \Mathematica{}
\cite{Wolfram:1999book,Wolfram:2005mathematica}, this requirement
amounts to using single words, strictly consisting of alphanumeric 
characters only. As a consequence, the user has to work
most of the time with an unconveniently labelled graphic 
and is furthermore required to keep track of the tags used in the 
substitution macros.

%\addtocounter{footnote}{1}
%\edef\remembermyfootnote{\number\value{footnote}} 
On the other hand, it is not always possible to use \Mathematica{}'s
conventional export function as it produces \acro{EPS} files requiring
the inclusion of additional fonts into the document, a process often
not being under the author's control.  A way
out is to include the font into the
\acro{EPS} file, or set the font family to a standard PostScript one:
\begin{alltt}\small
Plot[\dots, TextStyle\Rule\{FontFamily\Rule{}"Times"\}]\hfill\textrm{and}
Export[\dots, ConversionOptions\Rule{}
\hfill\{"IncludeSpecialFonts"\Rule{}True\}]\textrm{.}
\end{alltt}
\begin{figure}[t]
\vspace{4.5ex}
\subfigure[Conventional \Mathematica{} plot \emph{before} using \MathPSfrag]{
\includegraphics[width=0.97\linewidth]{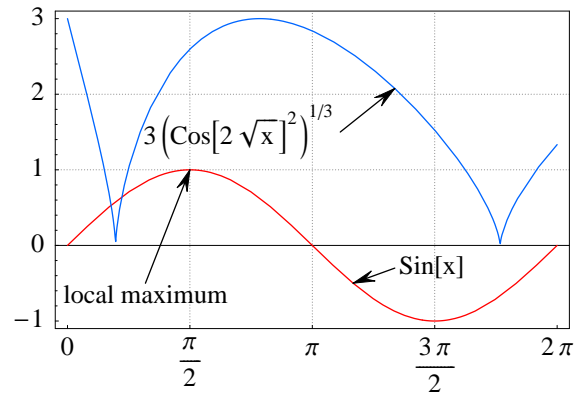}\label{fig:poor}
}
\begin{psfrags}
\subfigure[The same plot \emph{after} automatically substituting all \mc{Text} 
  primitives (including tick mark labels) by \LaTeX.]{
  \input{\graphicsdir ex_auto-psfrag.tex}
  \includegraphics[width=0.87\linewidth]{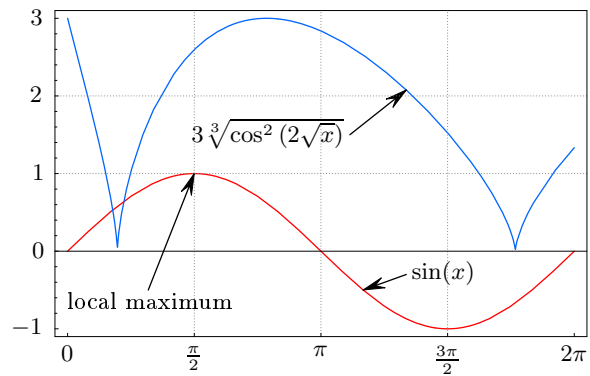}
  \label{fig:beautiful}
}  
\end{psfrags}
\caption{Old vs. new graphics export mechanism.}
\end{figure}
\begin{figure}[b]
%\vspace{1ex} 
\footnoterule\footnotesize
\begin{alltt}
\(\sp*\) jgrosse@mppmu.mpg.de
  \url{http://wwwth.mppmu.mpg.de/members/jgrosse}
\end{alltt}
\vspace{-3.5ex}
\end{figure}
\begin{figure*}
     \psfragdebugon
     \psfrag{gA}[br][br]{---[br][br]---}
     \psfrag*{gA}[Br][bc][2]{---[Br][bc][2]---}
     \psfrag*{gA}[cr][bl]{---[cr][bl]---}
     \psfrag*{gA}[tr][Bl]{---[tr][Bl]---}
     \psfrag*{gA}[bc][Bc]{---[bc][Bc]---}
     \psfrag*{gA}[Bc][Br]{---[Bc][Br]---}
     \psfrag*{gA}[cc][cr]{---[cc][cr]---}
     \psfrag*{gA}[tc][cc][0.75][45]{---[tc][cc][0.75][45]---}
     \psfrag*{gA}[bl][cl][1.5][30]{---[bl][cl][1.5][30]---}
     \psfrag*{gA}[Bl][tl]{---[Bl][tl]---}
     \psfrag*{gA}[bl][Bl]{~~~~~(baseline)}
     \psfrag*{gA}[bl][cl]{~~~~~(center line)}
     \psfrag*{gA}[bl][tc][1][-90]{~~~~~(center line)}
     \psfrag*{gA}[cl][tc]{---[cl][tc]---}
     \psfrag*{gA}[tl][tr][1][180]{---[tl][tr][1][180]---}
     \includegraphics[width=0.58\linewidth]{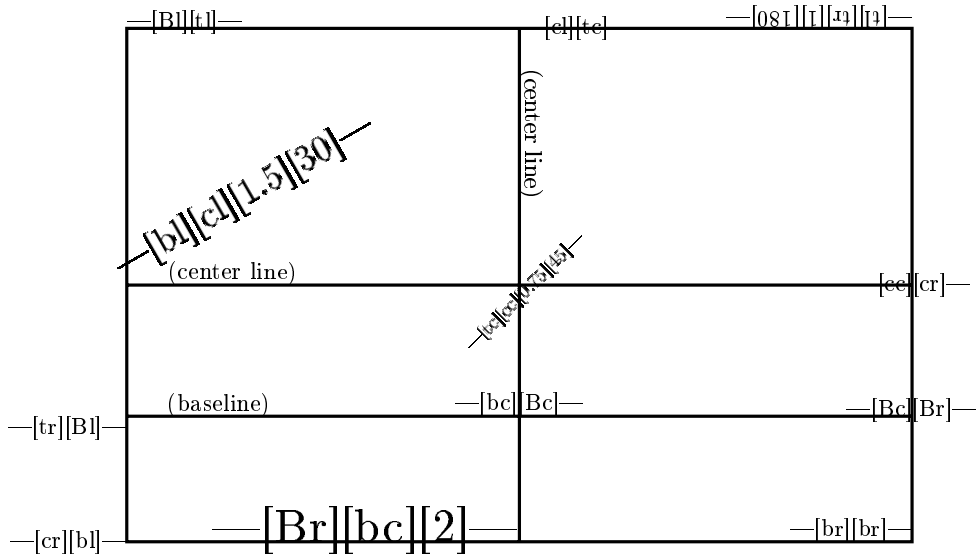}
\caption{Illustration of the various optional arguments of the
\cs{psfrag} command, taken from \cite{Grant:1998} with minor changes.
The first option determines the alignment of the \LaTeX\ description,
while the second one is responsible for the point to which the \LaTeX\
macro is attached.
\label{fig:psfragopts}}
\end{figure*}

While the slight mismatch between the PostScript font's appearance,
cf.~fig.~\ref{fig:poor}, and that of \LaTeX's standard font (Computer
Modern) may be acceptable in case of ordinary text labels,
mathematical expressions like square roots or fractions cannot compete
with \LaTeX's typesetting quality in this approach.  Consequently,
some authors simply restrict labelling of \Mathematica{} plots to a
bare minimum.

\MathPSfrag{} \cite{Grosse:2005} is a package that conveniently
produces publication-quality labels in \acro{EPS} files generated by
\Mathematica{}.  \MathPSfrag{} automates many (often all) tedious
details related to the use of the standard \LaTeX\ package \PSfrag{},
while still allowing manual fine tuning. As a demonstration of the
degree of automation, compare fig.~\ref{fig:poor}, which has been
generated by using the standard \Mathematica{} command \mc{Export},
and fig.~\ref{fig:beautiful}, generated by \MathPSfrag's export
instruction.

While the solution presented here, relies on the \PSfrag{} 
package, it avoids many of its shortcomings by providing
a semi-automatic layer. In particular,
\begin{itemize}
\item in most cases, it is sufficient to simply use the
      new \mc{PSfragExport} command,
\item including the graphic into the document requires
      only one additional command.
\end{itemize}

This article is organized as follows. A short review of \PSfrag{} and
a somewhat longer explanation of the semi-automatic features provided
by \MathPSfrag{}, are given in the first and second section,
respectively.  The third part contains several examples whose code as
well as that of figures \ref{fig:poor} and \ref{fig:beautiful} is
contained in the appendix.  Finally, some of \MathPSfrag's internals
are discussed.

\section{Review of \PSfrag{} \label{sec:psfragsty}}
This is intended to be a short introduction into \PSfrag{} explaining
only the essential features necessary to understand the corresponding
\Mathematica{} package's internals and to take advantage of its manual
options if automatic placement does not yield the desired result. The
full documentation can be found in \cite{Grant:1998}.

\PSfrag{} provides the macro
\begin{alltt}
\small\cs{psfrag}{\marg{tag}}[\oarg{posn}][\oarg{psposn}][\oarg{scale}][\oarg{rot}]{\marg{\textrm{\LaTeX}}}\textrm{,}
\end{alltt}
which replaces any occurence of \marg{tag} in the output of an
\acro{EPS} file by \marg{\textrm{\LaTeX}}.  According to
\cite{Grant:1998}, ``all \cs{psfrag} calls that precede an
\cs{includegraphics} (or equivalent) in the same or surrounding
environments'' will affect the output of the included graphics,
i.e.~\cs{psfrag} commands can be defined either locally, to act on stricly one
graphic, or globally, thus acting on all graphics in a
document.

\oarg{posn} and \oarg{psposn} are optional arguments which allow to
set (first) the vertical (top,
bottom, Baseline, or
center) and (second) the horizontal
(left, right, center) alignment of the replacement text by
specifying the respective first character of the choices given in
parentheses.  The arguments refer to the position of the reference
point in the respective bounding boxes.  The \LaTeX\ construct is
placed such that its reference point is at the position of the
corresponding PostScript (tag) box' reference point,
cf.~fig.~\ref{fig:psfragopts}.
\begin{figure}
\begin{center}
\begin{psfrags}
\input{\graphicsdir ex_rot-psfrag.tex}
\includegraphics[width=0.9\linewidth]{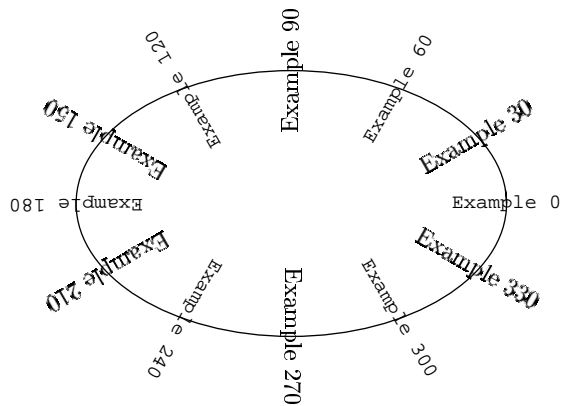}
\end{psfrags}
\end{center}
\caption{Example for substituting rotated text. To demonstrate that
the new export function can preserve the orientation, only half of
the labels have been substituted by \LaTeX. \label{fig:rot}}
\end{figure}

\oarg{scale} and \oarg{rot} permit scaling and rotation of the
inserted box, where the rotation is relative to the slope
of the PostScript bounding box such that a value 
of ``0'' preserves the orientation, see fig.~\ref{fig:rot}.  
Scaling is
best achieved by using \LaTeX\ scaling commands, like \cs{Large},
instead of the \oarg{scale} option, since the standard \LaTeX\ fonts 
consists of bitmaps rendered specifically for the chosen size and do not
rescale well. As will be demonstrated in the example section,
\MathPSfrag{} provides macro hooks that allow to scale labels 
retroactively from within the document.

\section{How to use \MathPSfrag{}}
There are only two commands needed to control \MathPSfrag's 
\acro{EPS} generation: \mc{PSfragExport}, which supersedes
\Mathematica's \mc{Export} command, and \mc{PSfrag}, which
allows overriding of the automatics for particular expressions.

The export function 
\begin{alltt}
PSfragExport[\marg{basename}, \marg{graphics}, \oarg{options}]
\end{alltt}
converts \marg{graphics}, the usual \mc{Graphics} construct returned
by \Mathematica{} commands like \mc{Plot}, to an \acro{EPS} file and a
\LaTeX\ file containing \cs{psfrag} macros.  

\oarg{options} can be any combination of the following options, 
listed with their parenthetic defaults.
\begin{itemize}
  \item \mc{TeXSuffix\Rule{}"\marg{string}"}\hfill(\mc{"-psfrag.tex"})
  \item \mc{EpsSuffix\Rule{}"\marg{string}"}\hfill(\mc{"-psfrag.eps"})
  \item \mc{RenumberTags\Rule{}\marg{boolean}}\hfill(\mc{False})
  \item \mc{AutoConvertText\Rule{}\marg{boolean}}\hfill(\mc{True})
  \item \mc{AutoPosition\Rule{}\marg{boolean}}\hfill(\mc{True})
\end{itemize}

The respective file names of the \LaTeX\ and \acro{EPS} file
are determined by \marg{basename} to which the value of the
options \mc{TeXSuffix} and \mc{EpsSuffix} is appended. 
\begin{figure}
\begin{psfrags}
\input{\graphicsdir ex_3d-psfrag.tex}
\includegraphics[width=0.8\linewidth]{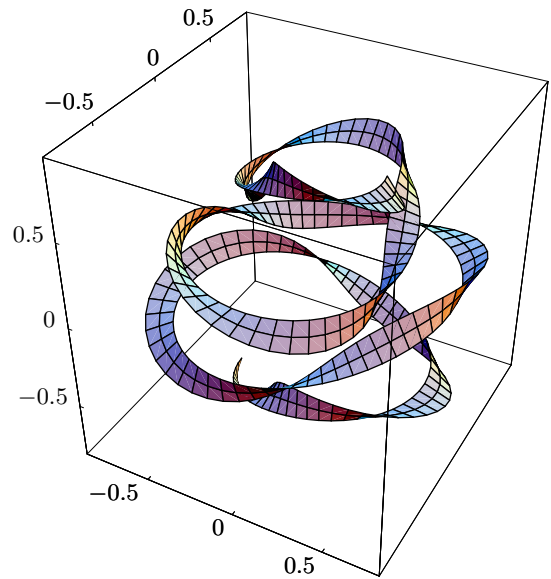}
\end{psfrags}
\caption{Three dimensional example: As there exists no
\mc{FullGraphics3D} command, manual labelling was required and the
\mc{RenumberTags} option of \mc{PSfragExport} was used to produce
smaller tags, increasing positioning precision.
\label{fig:three-d}}
\end{figure}

The option \mc{RenumberTags\Rule{}True} will renumber all tags and
represent the number as one of 52 (small and capital) 
latin characters or a combination of letters when the number is larger than
52. This feature, which generates very short tags, has been used in
fig.~\ref{fig:three-d} to achieve a preciser positioning.

Setting \mc{AutoConvertText\Rule{}False} will restrict conversion of
\mc{Text} directives in \marg{graphics} into \LaTeX\ commands to those
directives having been marked manually. The default behaviour is to
wrap the \mc{PSfrag} command discussed below around all expressions 
found in \marg{graphics}.

\mc{AutoPosition\Rule{}False} switches off the mechanism for 
determining the \cs{psfrag} alignment from \Mathematica's internal
representation of the graphics. Note that this also implies 
\mc{AutoConvertText\Rule{}False}.

Any other options will be passed on to \mc{Export} or applied to the
graphics using a \mc{Show} commmand, respectively.

For the purpose of manually controlling the output, that means circumventing
the automatics, the 
\begin{alltt}
PSfrag[\marg{expr}, \oarg{options}]
\end{alltt}
command is available.  It can be wrapped around each \Mathematica{}
expression \marg{expr} appearing as text in a graphic, such as the argument of a
\mc{PlotLabel\Rule\dots} or \mc{AxesLabel\Rule\dots} option. 

\mc{PSfrag} processes the following options, whose defaults have
been put in parentheses.
\begin{itemize}
  \item \mc{TeXCommand\Rule{}"\marg{string}"}\hfill(\mc{Automatic})
  \item \mc{PSfragTag\Rule{}"\marg{string}"}\hfill(\mc{Automatic})
  \item \mc{Position\Rule{}"\marg{yx}"}\hfill(\mc{Automatic})
  \item \mc{PSPosition\Rule{}"\marg{yx}"}\hfill(\mc{CopyPosition})
  \item \mc{Rotation\Rule{}"\marg{number}"}\hfill(0)
  \item \mc{Scaling\Rule{}"\marg{number}"}\hfill(Automatic)
\end{itemize}

Actually, \mc{PSfragExport}'s automatic mechanism simply
wraps \mc{PSfrag} around all \mc{Text} primitives using the 
default values above.
However, manual wrapping has the advantage
of allowing to apply different options to expressions where the
automatic behaviour did not give satisfactory results.

\mc{TeXCommand\Rule{}"\marg{string}"} sets to \marg{string} the \LaTeX\ command 
to appear in the final \acro{EPS} graphic as
a replacement of the corresponding expression \marg{expr}. If set to
\mc{Automatic}, the internal function \mc{GuessTeX} is called, which
is basically a wrapper around \mc{TeXForm} that adds {\$} signs around
math expressions. Moreover, \mc{GuessTeX} inserts some \LaTeX\ commands that can be used
to change the text style from within your document later on.

\mc{GuessTeX} has several options of the form \mc{PreApply\marg{type}\Rule{}\{\marg{list}\}}
and \mc{PostReplace\marg{type}\Rule{}\{\marg{list}\}}, where \marg{type} is one
of ``\mc{Text}'', ``\mc{Math}'' or ``\mc{Numeric}''. 
For each type respectively, they provide a
list for hooking in commands applied before or string replacements
applied after \mc{TeXForm}.  Especially,
the latter is rather useful for working around minor shortcomings of
earlier \Mathematica{} versions' \mc{TeXForm}.
\begin{figure}[t]
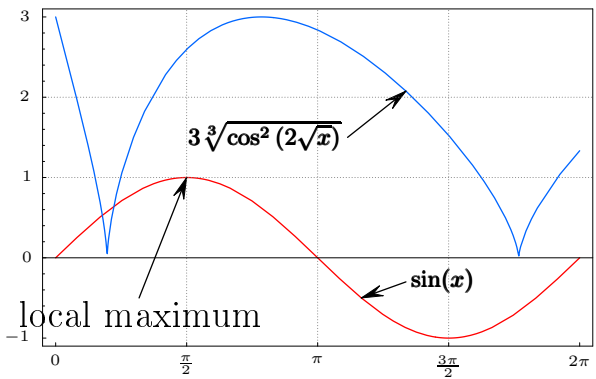

\begin{psfrags}
\newcommand{\psfragtextstyle}{\Large}
\newcommand{\psfragmathstyle}{\pmb}
\newcommand{\psfragnumericstyle}{\scriptstyle}
\input{\graphicsdir ex_auto-psfrag.tex}
\includegraphics[width=0.9\linewidth]{\graphicsdir ex_auto-psfrag.eps}
\end{psfrags}
\caption{Using the \LaTeX\ hooks included by \mc{GuessTeX} to fine-tune the appearance. As one can see, the
default values are chosen carefully.\\
\label{fig:terrible}}
\end{figure}

The remaining options are in one-to-one correspondence with those of
\cs{psfrag} explained in section \ref{sec:psfragsty}.  Their
respective default value \mc{Automatic} has the following different
meanings for each of them: For \mc{PSfragTag}, it means that a
\cs{psfrag} compatible tag is created from a string representation of
\marg{expr}, for \mc{Position} and \mc{PSPosition}, it means to take
over the alignment of the surrounding \mc{Text} command. (If there is
none, it waits till the \mc{Text} command is produced during export.)
For \mc{Rotation}, \mc{Automatic} means ``0'', whereas for
\mc{Scaling} it means insertion of one of three \LaTeX\ hooks,
\cs{psfragscaletext}, \cs{psfragscalemath} and \cs{psfragscalenumeric}
depending on the type of \marg{expr}.  The default value
\mc{CopyPosition} of \mc{PSPosition} does exactly what it says,
i.e.~taking over the value of the \mc{Position} option.

\subsection*{In the \LaTeX\ document}
There are only two additional things the user has to do:
\begin{enumerate}
\item add \cs{usepackage{psfrag, graphicx, amsmath}} into the
      document's preamble,
\item use \cs{input}\latex{\{\marg{basename}-psfrag.tex\}\}} to read the additional
      \cs{psfrag} labels created with \mc{PSfragExport["\marg{basename}", myplot]}.
\end{enumerate}

\section{Examples}
We start to consider in more detail figures \ref{fig:poor} and \ref{fig:beautiful}. 
The first one has been generated
using standard \Mathematica{} commands only, for the latter, the export
was carried out with \mc{PSfragExport["example", exampleplot]} and
it was included into this document with
{\small
\begin{verbatim}
\begin{psfrags}
\input{example-psfrag.tex}
\includegraphics[width=0.9\linewidth]
  {example-psfrag.eps}
\end{psfrags}.
\end{verbatim}
}
The \env{psfrags} starts an empty group provided by \PSfrag{}, whose 
sole purpose is making \cs{psfrag} definitions local to the
following graphic.

There are three \LaTeX\ commands, one of which is inserted by the
automatic \LaTeX\ guesser depending on the type of expression:
\cs{psfragtextstyle}, \cs{psfragmathstyle} and \cs{psfragnumericstyle}.
The latter is used for expressions identified by \mc{NumericQ}.  For
demonstration of their respective effects, the following lines
\begin{verbatim}
\newcommand{\psfragtextstyle}{\Large}
\newcommand{\psfragmathstyle}{\pmb}
\newcommand{\psfragnumericstyle}{\scriptstyle}
\end{verbatim}
have been inserted just before the \cs{input} command of
fig.~\ref{fig:terrible}.

\begin{figure}
\begin{psfrags}
\input{\graphicsdir ex_manual-psfrag.tex}
\includegraphics[width=0.9\linewidth]{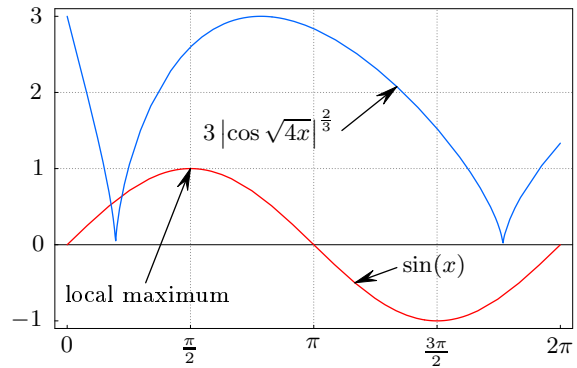}
\end{psfrags}
\caption{Example plot without resorting to automatics:
\mc{AutoConvertText\Rule{}False}, \mc{AutoPosition\Rule{}False}.
Additionally, 
the ``$\cos \dots$'' label's typesetting has been manually improved.
\label{fig:manual}}
\end{figure}

Fig.~\ref{fig:manual} demonstrates, that it is possible
to easily reconstruct fig.~\ref{fig:beautiful} without using
the automatic positioning feature.
Additionally,  one of the  labels' \LaTeX\ code was improved to be 
$3 \left|\cos \sqrt{4x}\right|$\raisebox{1ex}{$\scriptstyle\frac{2}{3}$} instead of 
$3 \sqrt[3]{\cos^2(2\sqrt{x})}$. 

\begin{figure}
\begin{psfrags}
\input{\graphicsdir ex_hold-psfrag.tex}
\includegraphics[width=0.9\linewidth]{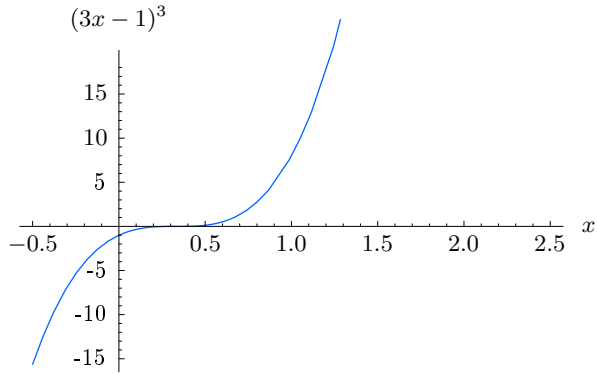}
\end{psfrags}
\caption{\mc{HoldForm} example: Without \mc{HoldForm}, \Mathematica{} would normal 
order the label on the $y$ axis to $(-1+3x)^3$. The \pack{CustomTicks} package
has been used to have a ``1.0'' instead of a ``1.'' on the $x$ axis.
\label{fig:hold}}
\end{figure}

Fig.~\ref{fig:hold} demonstrates compatibility with the \pack{CustomTicks}
package \cite{Caprio:2005} and the \mc{HoldForm} command, which
can be used to circumvent \Mathematica's automatic reordering of expressions
into a normal form. 

While \Mathematica{} does not reliably rotate text in an interactive session, 
\mc{PSfragExport} has no problems in doing so, as has
been shown in fig.~\ref{fig:rot}. Note that for each piece of text, 
the \mc{Rotation} option is set to ``0'', thus preserving the original 
orientation of the PostScript text.

Finally, it has been demonstrated in fig.~\ref{fig:three-d}, that
three dimensional graphics can be processed also, even though it has
to be done manually with \mc{PSfrag} commands, since the
\mc{FullGraphics} command only works on two dimensional graphics.

\section{Discussion\label{sec:discussion}}
\MathPSfrag{} relies on two \Mathematica{} commands: \mc{TeXForm} and
\mc{FullGraphics}. Both are potential sources of failure.

For the first one, this is due to substantial changes concerning its
output in its version history, which do not seem to have been 
publicly documented.
With the latest steps in the
transition towards \amsmath\ compatible \LaTeX\ though, 
\mc{TeXForm} will probably stabilize.  However,
\Mathematica{} versions 4.x and earlier will likely either require to include
additional style files shipped with these versions to process the
generated \LaTeX\ commands or to manually produce ordinary \LaTeX\
code. There are two possibilities to achieve the latter. First, one can
substitute all non-standard \LaTeX\ commands by setting up
\mc{GuessTeX}'s \mc{PostReplace\dots\Rule\{\dots\}} options accordingly. This
works particularly well if there is only a small number of non-standard
macros generated for a large number of text entries. Second, 
it is still possible to set all \mc{TeXCommand}s with
\mc{PSfrag}.

The automatic positioning relies on \mc{FullGraphics} to substitute
all \mc{Text} generating graphics options by \mc{Text} commands, which
in turn are used to read off the correct alignment for
\cs{psfrag}. However, as one can see comparing figures \ref{fig:poor} and
\ref{fig:beautiful}, the bounding box
differs slightly between \mc{Export[graphics]} and
\mc{Export[FullGraphics[graphics]]}.  
There might be further differences, which can not be corrected by
simply rescaling the graphics.
Therefore, \mc{PSfragExport} allows
to set \mc{AutoPosition\Rule{}False}, disabling the use
of \mc{FullGraphics}. In this case it has to use
static standard values when encountering an \mc{Automatic}
value, which cannot be interpreted anymore. (These fall back values
are: Bottomline, centered horizontally.)
Since the mechanism for converting options like \mc{PlotLabel} into 
\LaTeX\ labels also depends on \mc{FullGraphics}, setting 
\mc{AutoPosition\Rule{}False} implies \mc{AutoConvertText\Rule{}False}.

\section{Conclusion}
\MathPSfrag{} provides a convenient interface to \PSfrag{}
permitting the generation of high-quality labels in \Mathematica{}
graphics. While it automatizes all tedious aspects of \PSfrag{},
it still allows to seamlessly override all of its internal
assumptions.  Finally, \MathPSfrag{} does not provide methods 
to construct correct tick mark \emph{contents} as it is strictly
focussed on shape. As shown in fig.~\ref{fig:hold}, it 
does however integrate well with the
\pack{CustomTicks} package \cite{Caprio:2005}, which provides that 
functionality.

\section*{Acknowledgements}
%\begin{acknowledgments}
I am grateful to Riccardo Apreda and Robert Ei\-sen\-reich for helpful
comments and discussion.  
%\end{acknowledgments}

\appendix

\onecolumngrid

\section*{Figure Source Code}
We assume that \MathPSfrag{} and \pack{CustomTicks} are placed where
they can be found by \Mathematica. 
\pack{CustomTicks} is only needed for one of the examples.
\begin{alltt}\small
Needs["MathPSfrag`"];
Needs["CustomTicks`"];
<< Graphics`Arrow`;
SetDirectory["/tmp/"]; (* set according to your needs *)
\end{alltt}
\subsection{Automatic Example}
This example produces the conventional \Mathematica{} plot in fig.~\ref{fig:poor}. 
Merely in the last line, a \MathPSfrag{} command is invoked to produce fig.~\ref{fig:beautiful}.
\begin{alltt}\small
f1[x_] := Sin[x];
f2[x_] := 3*((Cos[2 Sqrt[x]])^2)^(1/3);

rawplot = Plot[\{f1[x], f2[x]\}, \{x, 0, 2 Pi\}, 
      PlotStyle\Rule{}\{Hue[1.0], Hue[0.6]\}, Frame\Rule{}True, 
      FrameTicks\Rule{}\{Pi/2*\{0, 1, 2, 3, 4\}, Automatic, None, None\}, 
      TextStyle\Rule{}\{FontFamily\Rule{}"Times"\}];

SimpleLabel[tip : \{_, _\}, txt_, txtpos : \{_, _\}, align : \{_, _\}] :=  Sequence[
    Arrow[txtpos, tip, HeadScaling\Rule{}Absolute, HeadLength\Rule{}8, HeadCenter\Rule{}0.6], 
    Text[txt, txtpos, align]];

textlabels = Graphics[\{
        SimpleLabel[\{Pi/2, f1[Pi/2]\}, "local maximum", \{1, -0.5\}, \{0, 1\}],
        SimpleLabel[\{7/6Pi, f1[7/6Pi]\}, f1[x], \{4.2, -0.3\}, \{-1, 0\}], 
        SimpleLabel[\{4.2, f2[4.2]\}, f2[x], \{3.5, 1.5\}, \{1, 0\}]
        \}];

mygrid = Map[\{\#, \{AbsoluteDashing[\{0.1, 1\}], GrayLevel[0.5]\}\} &, \{Pi*\{1/2, 1, 
            3/2\}, \{1, 2\}\}, \{2\}];
exampleplot = Show[rawplot, textlabels, GridLines\Rule{}mygrid];

Export["ex_nopsfrag.eps", exampleplot, "EPS"]
PSfragExport["ex_auto", exampleplot]
\end{alltt}

\subsection{Rotated Text}
While \Mathematica{} does not rotate the letters of a rotated
\mc{Text} on screen, both the conventional \mc{Export} and
\mc{PSfragExport} do the right thing, cf.~fig.~\ref{fig:rot}. 
Furthermore, it is demonstrated, that \mc{PSfragExport} can apply the
option \mc{PlotRange\Rule{}All} to the graphics before carrying out
the export.
\begin{alltt}\small
Show[Graphics[\{
        Table[Text["Example " <> ToString[Round[phi*180/Pi]], 
                   \{Cos[phi], Sin[phi]\}, \{0, 0\}, \{Cos[phi], Sin[phi]\}], 
              \{phi, 0, 2Pi - 0.01, 2Pi/6\}],
        Table[Text[PSfrag["Example " <> ToString[Round[phi*180/Pi]]], 
                   \{Cos[phi], Sin[phi]\}, \{0, 0\}, \{Cos[phi], Sin[phi]\}], 
              \{phi, 2Pi/12, 2Pi - 0.01, 2Pi/6\}], 
        Circle[\{0, 0\}, 1]
     \}]];
PSfragExport["ex_rot", \%, AutoConvertText \Rule{} False, PlotRange \Rule{} All]
\end{alltt}

\subsection{Three-dimensional Knot}
Here, the three-dimensional knot in fig.~\ref{fig:three-d} is generated.
Note that \mc{PSfragExport} acting on \mc{Graphics3D} always
implies \mc{AutoConvertText\Rule{}False} and \mc{AutoPosition\Rule{}False}.
\begin{alltt}\small
myticks3d = \{\#, PSfrag[\#, Position\Rule{}"Br"]\} & /@ \{-1, -0.5, 0, 1, 0.5, 1\};
ParametricPlot3D[
    Evaluate[Flatten[\{(0.5 + 0.2*Cos[phi/5] + r*Sin[1.7phi])\{Cos[phi], 
              Sin[phi]\}, phi/(5Pi) + r*Cos[1.7 phi]\}]],
    \{phi, -3Pi, 4Pi\}, \{r, 0.05, 0.2\}, PlotPoints\Rule{}\{200, 3\}, Ticks\Rule{}\{myticks3d, myticks3d, myticks3d\}];
PSfragExport["ex_3d", \%, RenumberTags\Rule{}True]
\end{alltt}

\subsection{Manual Clone of the Introductory Example}
Under the assumption that the automatic export did not work at all, 
the manual alignment options are used to reproduce fig.~\ref{fig:beautiful}.
Moreover, the opportunity to improve the $\cos(\dots)$ label by hand is seized; the corresponding
commands below are in italics. The functions
\mc{f1}, \mc{f2} and \mc{SimpleLabel} from the first example have been taken over.
\begin{alltt}\small
mytickmarks = \{
      \{N[\#], PSfrag[\#, Position\Rule{}"tc"]\} & /@ (Pi/2*\{0, 1, 2, 3, 4\}),
      \{N[\#], PSfrag[\#, Position\Rule{}"cr"]\} & /@ \{-1, 0, 1, 2, 3\}, 
      None, 
      None\};

\textit{texstr = "\$3\textbackslash{}left|\textbackslash{}cos \textbackslash{}sqrt\{4x\}\textbackslash{}right|^\textbackslash{}frac\{2\}\{3\}\$";}

Show[ Plot[
        \{f1[x], f2[x]\}, \{x, 0, 2 Pi\}, PlotStyle\Rule{}\{Hue[1.0], Hue[0.6]\}, 
        Frame\Rule{}True, FrameTicks\Rule{}mytickmarks, GridLines\Rule{}mygrid, 
        DisplayFunction\Rule{}Identity
        ],
      Graphics[\{
        SimpleLabel[\{Pi/2, f1[Pi/2]\}, 
          PSfrag["local maximum", Position\Rule{}"tc"], \{1, -0.5\}, \{0, 1\}],
        SimpleLabel[\{7/6Pi, f1[7/6Pi]\}, 
          PSfrag[f1[x], Position\Rule{}"cl"], \{4.2, -0.3\}, \{-1, 0\}], 
        SimpleLabel[\{4.2, f2[4.2]\}, 
          PSfrag[f2[x], Position\Rule{}"cr", \textit{TeXCommand\Rule{}texstr}], \{3.5, 1.5\}, \{1, 0\}]
        \}],
      DisplayFunction\Rule{}\$DisplayFunction
      ];
PSfragExport["ex_manual", \%, AutoConvertText\Rule{}False, AutoPosition\Rule{}False]
\end{alltt}

\subsection{HoldForm plus CustomTicks}
Fig.~\ref{fig:hold} is a plain example demonstrating that
\mc{HoldForm} can be used to fix the shape of an expression while
\mc{LinTicks} from \pack{CustomTicks} \cite{Caprio:2005} can be used
to circumvent the usual stripped decimal ``1.'' and print a much nicer
``1.0'' instead.
\begin{alltt}\small
Plot[(3x - 1)^3, \{x, -0.5, 2.5\}, PlotStyle\Rule{}Hue[0.6], 
    AxesLabel\Rule{}\{x, HoldForm[(3x - 1)^3]\}, 
    Ticks\Rule{}\{LinTicks[-0.5, 2.5], LinTicks[-15, 18]\}];
PSfragExport["ex_hold", \%]
\end{alltt}

\twocolumngrid
\bibliography{mathpsfrag}
\end{document}